\documentclass[12pt]{article}
\usepackage{pstricks-add,pst-slpe,times}

\usepackage{changepage}
\usepackage{amsmath,amssymb,amsthm,eurosym,lineno,mathrsfs}
\usepackage{graphicx}
\RequirePackage[colorlinks=true,allcolors=blue]{hyperref}

\usepackage{float}
\numberwithin{equation}{section}

\def\dd {\,{\rm d}}

\begin{document}


\title{\bf\large A Probabilistic Analysis of Autocallable Optimization Securities}

\author{{Gilna K. Samuel}\thanks{Department of Finance, 
Rensselaer Polytechnic Institute, Troy, NY 12180.} 
\ \ {and \ Donald St. P. Richards}\thanks{Department of Statistics, 
Pennsylvania State University, University Park, PA 16802.
\endgraf
\ {\it 2000 Mathematics Subject Classification}: Primary 60E99.
\endgraf
\ {\it Key words and phrases}.   Financial derivatives; Law of Total Expectation;  Return optimization securities; Contingent Protection; S\&P 500 Financials Index.}
}

\date{September 6, 2016}

\maketitle

\begin{abstract}
\footnotesize{
We consider in this paper some structured financial products, known as reverse convertible notes, that resulted in substantial losses to certain buyers of these notes in recent years. We shall focus on specific reverse convertible notes known as ``Autocallable Optimization Securities with Contingent Protection Linked to the S\&P 500 Financial Index,'' because these notes are representative of the broad spectrum of reverse convertibles notes. Therefore, the analysis provided in this paper is applicable to many other reverse convertible notes.  

We begin by describing the notes in detail and identifying potential areas of confusion in the pricing supplement to the prospectus for the notes. We deduce two possible interpretations of the payment procedure for the notes and apply the Law of Total Expectation to develop a probabilistic analysis for each interpretation. We also determine the corresponding expected net payments to note-holders under various scenarios for the financial markets and show that, under a broad range of scenarios, note-holders were likely to suffer substantial losses.

As a consequence, we infer that the prospectus is sufficiently complex that financial advisers generally lacked the mathematical knowledge and expertise to understand the prospectus completely. Therefore, financial advisers who recommended purchases of the notes did not have the knowledge and expertise that is required by a fiduciary relationship, hence were unable to exercise fiduciary duty, and ultimately misguided their clients. We conclude that these reverse convertibles notes were designed by financial institutions to insure themselves, against significant declines in the equities markets, at the expense of note-holders.  
}
\end{abstract}

\parskip=2.5pt

\section{Introduction}
\label{intro}
\setcounter{equation}{0}

In July, 2011 the U.S. Financial Industry Regulatory Authority (FINRA) issued an alert \cite{finra} concerning the sale of structured financial products known as ``reverse convertible notes.'' These notes were issued by many financial firms and sold widely to clients who had an optimistic view of future market conditions and who were seeking to diversify their financial portfolios. Clients were promised high yields, and phrases such as  ``autocallable optimization''  or ``contingent protection'' in the titles of the notes imbued clients with a high sense of confidence in the future of these notes.  However, the notes turned out not to be as simple or as safe as customers had thought initially, and buyers of these notes eventually suffered enormous losses when the financial markets experienced a significant downturn in 2008.

The U.S. Securities Exchange and Commission (SEC) defines a reverse convertible note to be a financial product whose return is linked to the performance of a ``reference asset" or a ``basket of reference assets,"  usually consisting of the stock price of an unrelated company or the level of a stock market index. Some common reference assets for such notes are the S\&P 500 Financials Index, the EURO STOXX 50, and the common stock of JPMorgan Chase \& Co. or other prominent corporations.  In this paper, we will investigate an autocallable note, a common type of reverse convertible note. Autocallable notes have one or more ``call dates," which are dates on which the note can be redeemed prior to the maturity date and which are determined by specific market conditions. A wide variety of reverse convertible notes have been sold to the public since 2006, with sales continuing even today, and these notes can be examined in a manner similar to the analysis given in this paper. Therefore, the analysis provided in this paper serves as a template for the study of many reverse convertibles notes.

In this paper, we investigate a particular reverse convertible note known as ``Autocallable Optimization Securities with Contingent Protection Linked to the S\&P 500 Financials Index."  This financial product was issued by Lehman Brothers Holdings Inc. in early 2008, before the 2007-2008 financial crisis reached its nadir, and the pricing supplement to the prospectus for this note can be obtained from the SEC's website at \cite{Lehman08}.  We will study the pricing supplement to the prospectus, and provide a description of the note and its key features. This pricing supplement is similar to the pricing supplements for many other reverse convertible notes issued by other financial companies. 

At first glance, the pricing supplement to the prospectus appears to provide a promising description of the note; however, on examining the pricing supplement in detail, we found some disturbing attributes. First, the term ``contingent protection'' in the title causes this financial product to appear more secure than it really is. The phrase ``contingent protection'' may suggest that there is some legal or financial protection on the notes, thereby causing some customers to become more confident about the returns on their purchases. Therefore, we will first assess the weaknesses of the contingent-protection feature of these notes. 

Another concern is that the details of the pricing supplement could have been confusing to a financially unsophisticated client and as a result, many clients would have found it difficult to understand complete details of the note. Our second task, therefore, is to describe some vaguenesses that we have found in the prospectus regarding the payment procedure and redemption rules. In order to assess these details we will analyze the method used to determine payment and describe two interpretations of the payment procedure. 

Next, we provide a probabilistic analysis of the payment procedure and use the Law of Total Expectation to calculate the expected net returns to clients. We will show, based on the results derived from our analysis, that customers were at a substantial disadvantage from the very moment they purchased this note. We deduce from our analysis that the average client would have lost a significant proportion of their principal; in some situations, those losses could be as high as 50\%. 

At this point, we provide some remarks on the concept of {\emph{fiduciary duty}}. Many state laws and the U.S. Investment Advisers Act of 1940 require that certain financial advisers act as {\emph{fiduciaries}} in their business transactions with clients. {\sl West's Encyclopedia of American Law} \cite{west08} defines a {\emph{fiduciary relationship}} as ``one which encompasses the idea of faith and confidence and is generally established only when the confidence is given by one and accepted by another.'' ``A fiduciary has greater knowledge and expertise of the matters being handled and is held to a standard of conduct and trust above that of a casual business person'' (Hill \cite{Hill02}). Fiduciary duty is  ``the obligation to act in the best interest of the beneficiary of the fiduciary relationship'' (Rahaim \cite{Rahaim05}).

As a consequence, we will conclude that financial advisers who recommended purchases of reverse convertible notes in early 2008 did not exercise fiduciary duty to their clients. Indeed, the prospectus is sufficiently complex that financial advisers themselves could not have understood every detail of the prospectus. Further, financial advisers lacked the mathematical knowledge and expertise necessary to determine the consequences of details in the prospectus. Therefore, financial advisers simply were unable to exercise fiduciary duty to their clients, and ultimately the advisers misguided their clients.

We remark that related probabilistic analyses of other types of financial derivatives were given by Richards and Hundal \cite{Richards1} and Richards \cite{Richards2}; cf. Samuel \cite{Samuel}. Those papers analyzed structured products, such as, return optimization securities, yield magnet notes, reverse exchangeable securities, and principal-protected notes. In each case, it was shown that purchasing these structured products during the mid-2000's would likely lead to substantial losses. In this paper, a similar probabilistic analysis will be done on reverse convertible notes and our analysis will show that the purchase of these notes were likely to result in significant losses also. 

Moreover,  Richards \cite{Richards2} observed that certain structured financial products were designed ``to insure the banks against substantial declines in the markets; such an arrangement allowed the banks to avoid direct stock sales on the open market, which could have triggered widespread market declines.'' Likewise, the reverse convertible notes analyzed in this paper provided a similar safety net to the financial institutions which issued them, but at the expense of note-holders. 

We conclude the introduction with a summary of the results to follow in the remaining sections of this paper.  In Section \ref{pricingsupplement}, we provide an assessment of the pricing supplement to the prospectus.  We will point out that seemingly positive features of the payment procedure for the notes appear to provide overly optimistic views of future market conditions seem to suggest that clients' principal are protected against adverse market conditions, and that the chances of a loss of principal are low.  Moreover, the pricing supplement presents positive features of the notes in the earlier part of the document, and in greater detail.  On the other hand, negative aspects of the notes are described in a less prominent manner; those aspects are relegated to later sections which are less likely to be read by a typical client; and the supplement does not provide a comprehensive list of possible payout scenarios for the notes.

We shall also identify problems with the payment procedure for the notes.  Indeed, the description of the payment procedure provided in the payment supplement seems incomplete and vague.  Moreover, the conditions under which the note will be called is unclear, leading to at least two plausible interpretations of the payment procedure.

In Section \ref{sec:firstinterpretation}, we describe and analyze a first interpretation of the payment procedure.  We shall use the Law of Total Expectation to obtain mathematical formulas for the expected, or average, payout to a randomly chosen note-holder under a variety of market-outlook scenarios ranging from pessimistic to optimistic.  By analyzing graphs of the expected net payment functions, we shall prove that for the expected net payment to be positive, the market outlook must necessarily be highly optimistic, and even in such a best-case scenario the note-holder is likely to be at a disadvantage to the issuing bank because of the call provisions of the note.  As regards pessimistic scenarios, we show that the average return to note-holders can be as low as -50\% in some instances.

In Section \ref{sec:secondinterpretation}, we describe a second plausible interpretation of the payment procedure.  We shall show that the expected net payment under this interpretation can never be greater than the expected net payment under the interpretation analyzed in Section \ref{sec:firstinterpretation}. Therefore, the outcomes for note-holders can be expected generally to be worse than those described in Section \ref{sec:firstinterpretation}.  In Section \ref{conclusions}, we present some closing remarks on our analysis.  We will conclude that these reverse convertible notes were simply clever marketing scheme used by financial institutions wanting to insure themselves against future market declines.

\section{Assessment of the Pricing Supplement to the Prospectus}
\label{pricingsupplement}
\setcounter{equation}{0}

Reverse convertibles notes known as ``Autocallable Optimization Securities with Contingent Protection Linked to the S\&P 500 Financials Index" were sold to the public in multiples of \$10. The pricing supplement \cite{Lehman08} to the prospectus states that the trade date for this financial product was February 5, 2008, with a settlement date three days later on February 8, 2008. These notes had an 18-month ``observation period,'' from May 5, 2008 to August 5, 2009, with a total of six ``observation dates'' every three months. The final valuation date was on August 5, 2009, and the maturity date was five days later on August 10, 2009. 

The pricing supplement also mentions information such as the advantages and disadvantages of the note and possible returns that clients might receive from their purchases. However, the information provided seem biased toward positive aspects of the note. For instance, the positive aspects are emphasized in a ``Features Section'' on the first page of the pricing supplement, and are highlighted in boldfaced font while the explanation of each feature is in a smaller font. The first feature listed states that the notes provide ``positive call return in flat or bullish scenarios,''  implying that clients will receive positive returns as long as the S\&P 500 Financials Index remains above a certain level.

Another positive attribute listed in the ``Features Section'' is that of ``contingent principal protection.'' This feature suggests that clients have some protection against loss of principal; however, the specified market conditions under which the contingent protection applies is printed in a smaller font on the same page.

The last feature states that the notes ``express a bullish view of the U.S. Financial Services Sector.'' Given that the notes are linked to the S\&P 500 Financials Index which comprises of 93 companies in the financial services sector of the S\&P 500 Index, financial advisers recommending purchases of these notes should have provided their clients with an estimate of the likelihood that the S\&P 500 Financials Index would continue on an upward trend. However, neither advisers nor the pricing supplement appear to have provided any such information.  

Indeed, the features listed in the ``Features Section''  seem to promote optimistic views of future market conditions, leading clients to believe that their future gains will be high. These details imply that the clients' principals are protected against adverse market conditions and their chances of a loss of principal are low. Moreover, buyers of these notes were more likely to read the first few sections in greater detail compared to the later sections. Thus, the positive aspects of these notes may have played a more influential role in clients' decisions to purchase the notes.

On the other hand, negative aspects of the note are portrayed in a less prominent manner and are postponed to later sections of the pricing supplement. For instance, the fact that the note is not insured by the Federal Deposit Insurance Corporation (FDIC) is stated in smaller font. The section ``Key Risks,'' which lists numerous risk factors, appears as one of the last sections in the pricing supplement. Also, the pricing supplement provides no mention of the likelihood that downward trends in the market can occur, leading to a total loss of capital. Moreover, the prospectus implies that future market conditions will be positive but then warns later against the expectation of a ``positive-return environment.''  Further, based on the pricing supplement's guidelines for net payment, the greatest possible return on the note is only 31.26\%, whereas the greatest possible loss is 100\% of capital. That is, profits on the note are limited to a specific percentage while there is no similar limit on the percentage of capital that clients might lose. 

Another major concern is the description of the payment method given in the pricing supplement; in our view the description is incomplete. The pricing supplement provides only four examples of possible scenarios of the future market trends.  In our view, the supplement should have provided a more comprehensive list of such scenarios. In two of the scenarios presented the client receives a positive return; in  one scenario the client breaks even, and in only one scenario does the client lose part of their principal. Also, no example is given to demonstrate how a client could have suffered a loss of their entire capital. Therefore, these examples may have led some clients to believe that they were more likely to receive positive returns rather than losses.

Moreover, the details in the scenario analysis are vague. Specifically, the determination of the conditions under which the note will be called is unclear, and therefore we will now investigate the payment procedure in great detail.

In order to describe the payment method, we need to define several terms. The {\emph{Index Starting Level}} is the closing level of the S\&P 500 Financials Index on February 5, 2008. The pricing supplement states the  Index Starting Level to be $369.44.$ The {\emph{Trigger Level}} is $184.72,$ which is 50\% of the Index Starting Level. The {\emph{Index Ending Level}} is the closing level of the S\&P 500 Financials Index on the corresponding observation or trade date. Having defined these terms, we can define the {\emph{Index Return}}, denoted by $I$, to be 
$$
I= \frac{\hbox{Index Ending Level} - \hbox{Index Starting Level}}{\hbox{Index Starting Level}}.
$$
By means of this formula, we see that if the Index Ending Level on any trade date is at or above the Index Starting Level then the Index Return will be positive, i.e. $I \geq 0$, while if on any trade date the Index Ending Level is less than the Index Starting Level then the Index Return will be negative, i.e. $I<0$.  Also, if the Index Ending Level is less than the Trigger Level then $I< -0.5$ and conversely.

The section of the pricing supplement entitled ``Payment at Maturity'' contains a diagram which is used to illustrate the rules for calculating net payments to clients. The diagram presented here is typical, and hence representative of related diagrams appearing in the pricing supplements of many other autocallable notes found via the SEC's website. This diagram is the primary source in the pricing supplement devoted to determining the payment procedure and it is reproduced in Figure \ref{payment}, as follows:

\begin{figure}[H]
\includegraphics[scale=0.9]{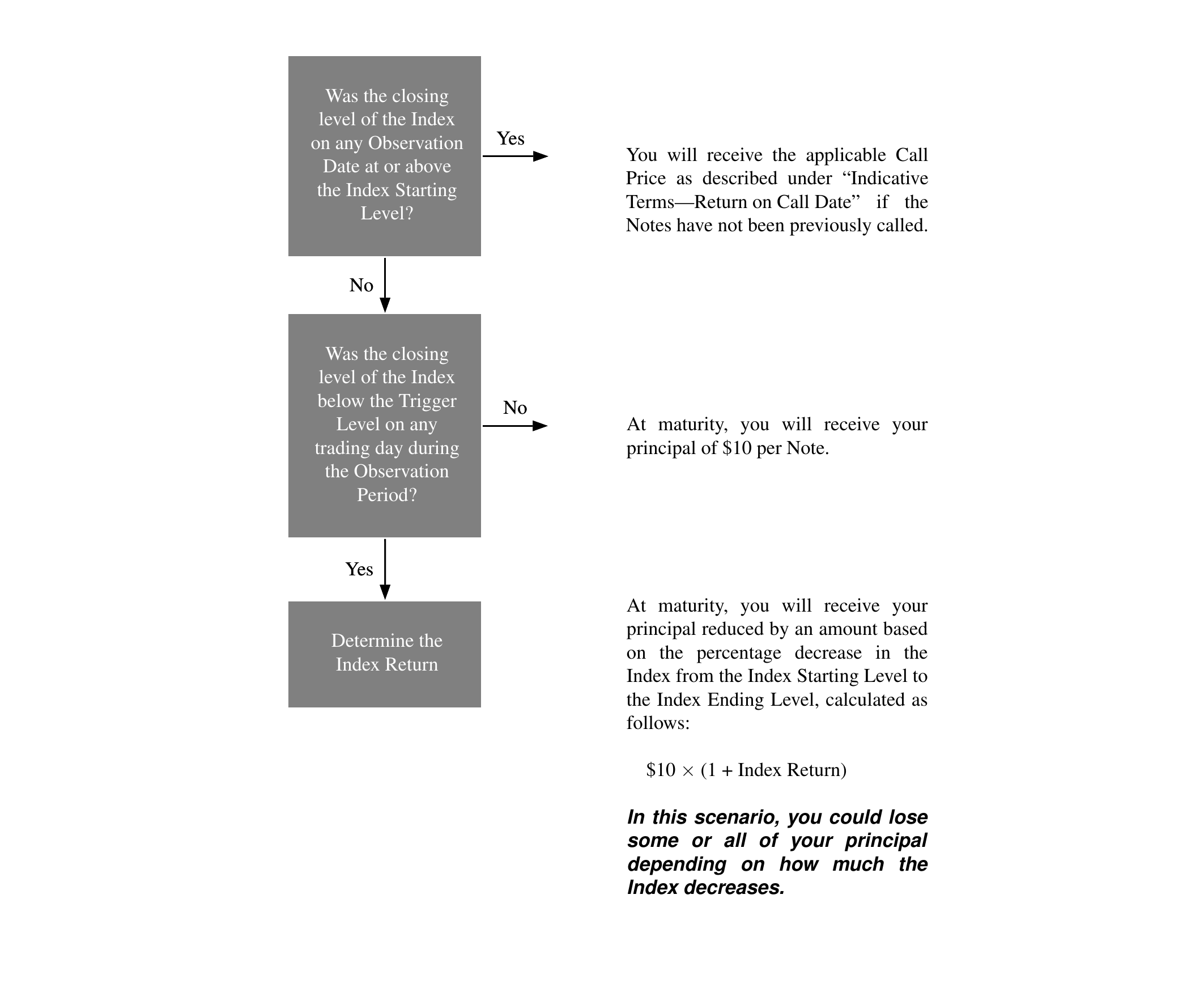}
\vskip-25pt
\caption{Payment at Maturity, as Depicted in the Pricing Supplement}
\label{payment}
\end{figure}


From Figure \ref{payment}, we deduce that if $I >0$ on an observation date, then the note {\emph{might}} be called and redeemed for an amount stated in the pricing supplement. This redemption value is based on a rate of 20.84\% per annum, with actual amounts stated in the ``Final Terms'' section of the supplement. This ``Final Terms'' section states each observation date and the corresponding percentage return if the note is called on that date.

However, if $I <0$ on a particular observation date then the note is not called. Also, if  $I<-0.5$ on any trading date during the observation period then there is a possibility that the client will receive a negative return at maturity. In particular, the diagram stated that the client can lose up to a 100\% of their capital but provided no indication of the probability of this occurrence. Also, the diagram itself does not make it explicit when the note will be called, but as long as the notes are not called then the calculation of the Index Return will be done on every subsequent trading date until the final valuation date. 

To determine the actual return that note-holders will receive on their notes is particularly vague. It appears that the notes could have been called before the final valuation date, and at the complete discretion of the sellers. Suppose also that the Index Return, when measured on the final valuation date, is substantially greater than zero, i.e., the Index Ending Level is higher than the Index Starting Level; then the language used in the pricing supplement appears to give the seller of the note the power to have called the note in a way so as to minimize the return to note holders. Therefore, it seems that the manner in which payments to clients are determined can be interpreted in different ways. In the following two sections we will analyze two likely interpretations of the payment procedure that could have been used and we will estimate the average return to clients under each interpretation.

\section{ A First Interpretation of the Payment Procedure}
\label{sec:firstinterpretation}
\setcounter{equation}{0}

In this section, we describe and analyze one of the possible interpretations of the payment procedure for calculating net payment. For this interpretation, we will show that even under optimistic scenarios, an average buyer will lose a substantial amount of their principal.  

For each observation date within the 18-month period from February 5, 2008 to August 5, 2009, we define the corresponding index return as follows:\\

$I_1$ : Index return on May 5, 2008

$I_2$ : Index return on August 5, 2008

$I_3$ : Index return on November 5, 2008

$I_4$ : Index return on  February 5, 2009

$I_5$ : Index return on  May 5, 2009

$I_6$ : Index return on  August 5, 2009\\

Let $n$ denote the total number of trading days during the 18-month observation period. We have verified through Google Finance's website on historical closing prices of the S\&P 500 Financials Index, that for this note $n=381$ trading days. For $1,\ldots, n$, define $d_i$ to be the cumulative index return for Day $i$; thus $d_i$ represents the total index return from Day 1, February 5, 2008, to Day $i$.

Let $d_{\min} = \min{(d_1,d_2,...,d_n)}$ be the smallest daily cumulative return over the entire observation period.

Using this notation, the steps for this interpretation of the payment procedure can now be described as follows:

\begin{itemize}
\item[ ] Step 1:\vspace{-7mm} \begin{adjustwidth}{1.5cm}{1.5cm} Calculate the Index Return, $I_r$  on the observation date for the $r$th quarter.
\end{adjustwidth}
\item[ ] Step 2: \vspace{-7mm} \begin{adjustwidth}{1.5cm}{1.5cm} If the Index Return on the observation date is non-negative, i.e. if $I_r\geq 0$, then the note is called and the client receives an amount as stated in the pricing supplement on the Final Valuation date.
\end{adjustwidth}
\item[ ] Step 3: \vspace{-7mm} \begin{adjustwidth}{1.5cm}{1.5cm} If the Index Return on the observation date is negative, i.e. if $I_r < 0$ then we calculate the cumulative returns, $d_i$, for each trading date during the quarter corresponding to the given observation date.  If any of those cumulative returns are less than $-50 \%$ then we record this occurrence for future reference.
\end{adjustwidth}
\item[ ] Step 4: \vspace{-7mm} \begin{adjustwidth}{1.5cm}{1.5cm} Return to Step 1 for the next observation date and repeat the process.\\
\end{adjustwidth}
\end{itemize}
Steps 1-4 are carried out on every observation date until the final valuation date or until the note is called. 

At the Final Valuation date, one of the following three cases can occur:

\begin{itemize}
\item[ ] Case 1:\vspace{-7mm} \begin{adjustwidth}{1.5cm}{1.5cm}   If the Index Return is non-negative on the Final Valuation date  i.e. if $I_6 \geq 0$, then the client receives the amount stated in the pricing supplement. 
\end{adjustwidth}
\item[ ] Case 2: \vspace{-7mm} \begin{adjustwidth}{1.5cm}{1.5cm} If the Index Return on the Final Valuation date is negative and every daily Index Ending Level remained at or above the Trigger Level on each trading date during the entire observation period, i.e. if $I_6 <0 $ and  $ d_{\min}\ge - 50\% $, then the client receives \$10 at maturity.
\end{adjustwidth}
\item[ ] Case 3:\vspace{-7mm} \begin{adjustwidth}{1.5cm}{1.5cm}  If on any trading date prior to the Final Valuation Date at least one daily Index Ending Level breached the Trigger Level, i.e. if $d_{\min} < -50\%$, then the client receives a reduced payment equal to $ 10(1+I_6)$, representing a negative return of  $I_6$ to note-holders.
\end{adjustwidth}
\end{itemize}
The following tree diagram describes this interpretation of the payment procedure:

\begin{figure}[H]
\hspace{-2.0in}
\vspace{-0.5in}
\includegraphics{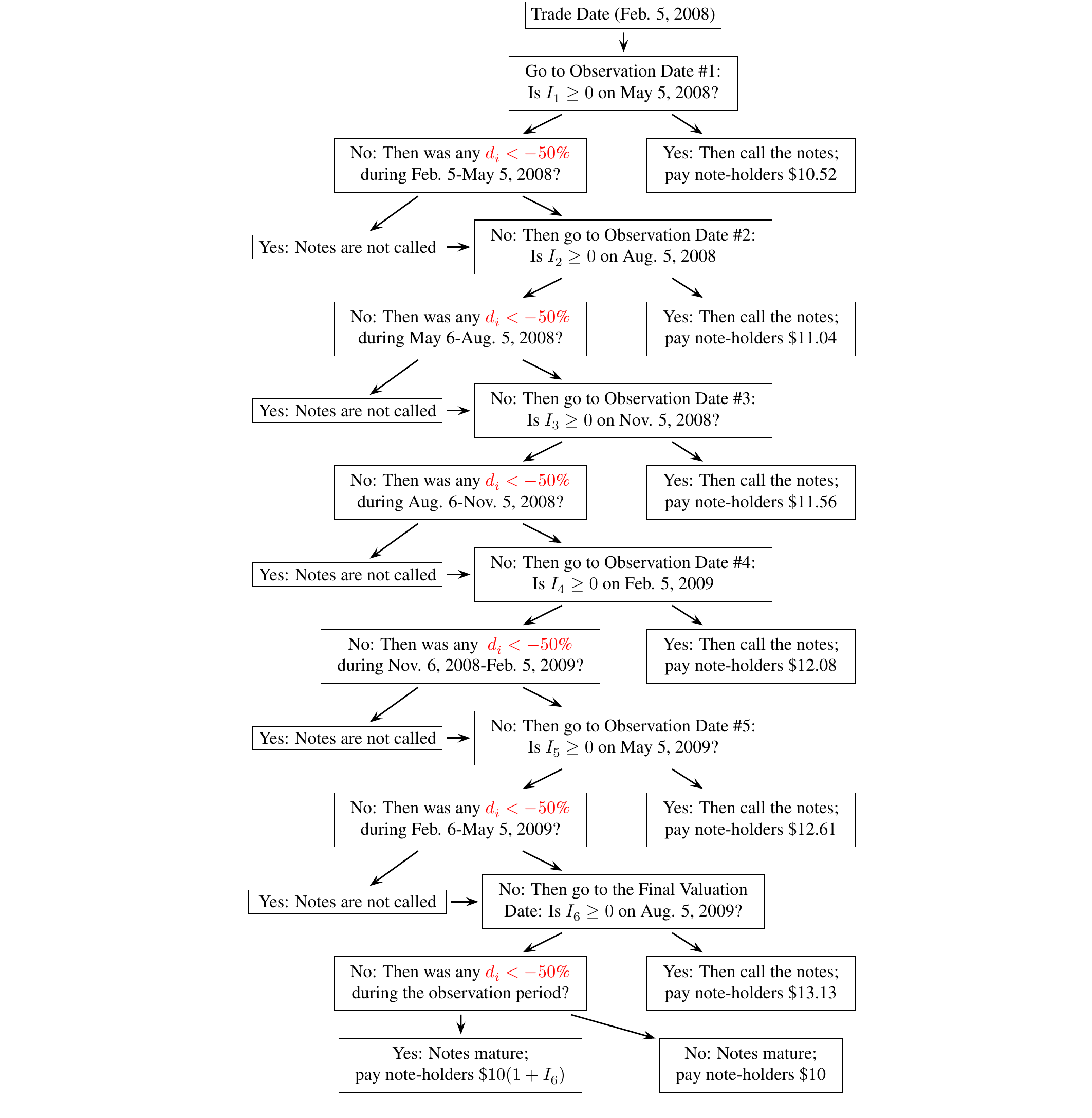}\\
\vspace{0.4cm}
\caption{Schematic Description of the Payment Procedure}
\label{payment1}
\end{figure}


 
Then, according to the pricing supplement \cite{Lehman08} and this payment procedure, the net payment in U.S. dollars, to note-holders is given by the function, 
\begin{equation} 
\label{netpayment1}
\hbox{Net Payment} = 
\begin{cases}
0.52, \hbox{ if } I_1 \ge 0 \\
1.04, \hbox{ if } I_1< 0, \hbox{ and }  I_2 \geq 0\\
1.56,  \hbox{ if } I_j < 0, j=1,2 \hbox{ and }  I_3 \geq 0 \\
2.08, \hbox{ if } I_j < 0, j=1,2,3 \hbox{ and }  I_4 \geq 0 \\
2.61, \hbox{ if } I_j < 0, 1 \leq j \leq 4 \hbox{ and }  I_5 \geq 0\\
3.13, \hbox{ if } I_j < 0, 1 \leq j \leq 5 \hbox{ and }  I_6 \geq 0\\
0,  \hspace{5mm} \hbox{ if } I_j < 0, 1 \leq j \leq 6 \hbox{ and } d_{\min} \geq -\tfrac 12\\
10 I_6,   \hbox{ if } I_j < 0, 1 \leq j \leq 6 \hbox{ and }  d_{\min} < -\tfrac 12
\end{cases} 
\end{equation}
\ 
\begin{itemize}
\item The first six cases occur if any of the index returns, $I_1, \ldots, I_6$ are non-negative, in which case, the note is called on the corresponding observation date, and the client receives a positive return on the Final Valuation date in those instances.
\item The seventh case occurs if  all of the index returns, $I_1, \ldots, I_6$ are negative and the minimum cumulative daily index return  during the observation period, $d_{\min}$ is at least $-50\%$. In this case, the note matures and the client receives a net payment of zero on the Final Valuation date, i.e. the client  breaks even.
\item The eighth and last case occurs if the note is not called previously and $d_{\min}$, {\it the minimum cumulative daily index return during the observation period}, is less than $-50\%$; i.e. the smallest Index Ending Level falls below the Trigger Level. In this case, the note matures and on the Final Valuation date the client receives a negative net payment of  $10I_6$.
\end{itemize}

In the net payment function in (\ref{netpayment1}), the index returns $I_1,\ldots,I_6$ are continuous random variables whose joint probability density function is strictly positive for $I_1,\ldots,I_6$ in the interval $(-1,0)$; i.e. there is a non-zero probability of a downward market trend in the future. Hence, 
\begin{equation}
\label{Prob(B)}
P(d_{\min} <-\tfrac 12|I_j < 0,  1 \leq j \leq 6) \neq 0 
\end{equation}
since there always exists the possibility of decreasing market trends. 

We now apply the Law of Total Expectation (Ross \cite{Ross10}, page 333, Equation (5.1b)) to the net payment function in (\ref{netpayment1}); then we obtain
\begin{equation}
\label{e(netpayment)}
\begin{aligned}
E(\hbox{Net Payment}) = \ & 0.52 \times P( I_1 \ge 0)\\
                         & + 1.04\times P(I_1< 0, \hbox{ and }  I_2 \geq 0) \\
       & + 1.56 \times P(I_j < 0, j=1,2 \hbox{ and }  I_3 \geq 0  ) \\
       & + 2.08 \times P(I_j < 0, j=1,2,3 \hbox{ and }  I_4 \geq 0  )\\
       & + 2.61 \times P(I_j < 0, 1 \leq j \leq 4\hbox{ and }  I_5 \geq 0     )\\
       & + 3.13 \times P(I_j < 0, 1 \leq j \leq 5 \hbox{ and }  I_6 \geq 0  )\\
  &   + 0 \times P(I_j < 0, 1 \leq j \leq 6 \hbox{ and }  d_{\min}\geq -\tfrac 12) \\
  & + 10\times E(I_6|d_{\min} <-\tfrac 12 \hbox{ and } I_j < 0,  1 \leq j \leq 6)\\
  & \quad \quad \times P(d_{\min} <-\tfrac 12|I_j < 0,  1 \leq j \leq 6) \\
  & \quad \quad \times P(I_j < 0, 1 \leq j \leq 6).
\end{aligned}
\end{equation}

In a real-world setting, the index return random variables $I_1,\ldots,I_6$ are unlikely to be independent. In fact, there are numerous research articles which have found evidence that momentum-trading ``strategies which buy stocks that have performed well in the past and sell stocks that have performed poorly in the past generate significant positive returns over 3- to 12-month holding periods'' (Jegadeesh and Titman \cite{Jegadeesh93}; Conrad and Kaul \cite{Kaul98}); this phenomenon has been observed to hold not only for U.S. financial markets but also internationally (Hurn \cite{Hurn03}). Therefore, if the financial index has recorded a negative return in a three-month period then it is natural to expect traders to be pessimistic about the ensuing three-month period. Therefore, we would expect to find that, for example,
\begin{equation}
\label{realworld}
 P(I_2 \ge 0 | I_1 < 0) \le P(I_2 < 0 | I_1 < 0) 
\end{equation}
in a real-world setting.  

We recognize that there can be exceptions to this inequality; if, for instance, the markets are in a euphoric state of mind, where any downturn is regarded optimistically as a sign of positive future developments then the above inequality might even be reversed at times.  Since it is unrealistic and unwise to treat the markets as being in a permanent state of euphoria, we will treat this inequality as being valid generally. Therefore for the purposes of analyzing the expected net payment to note-holders, we will assume that the inequality (\ref{realworld}) holds.  

We remark that because 
\begin{equation}
\label{sumrealworld}
P(I_2 \ge 0 | I_1 < 0) + P(I_2 < 0 | I_1 < 0) =1,
\end{equation}
then the inequality (\ref{realworld}) implies that
\begin{equation}
\label{realworld2}
P(I_2 \ge 0 | I_1 < 0) \le \tfrac 12 \le P(I_2 < 0 | I_1 < 0).
\end{equation}
Also, (\ref{sumrealworld}) implies that if $P(I_2 \ge 0 | I_1 < 0) $ is large, i.e. close to one, then, $P(I_2 < 0 | I_1 < 0)$ is small, i.e. close to zero.

\begin{figure}[H]
\centering
\includegraphics[scale=0.7]{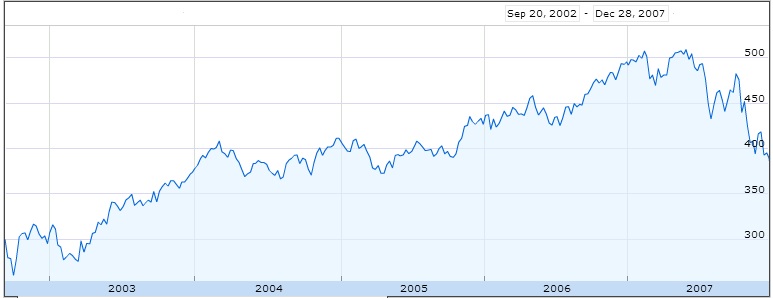}
\caption{Closing Levels of the S\&P 500 Financials Index for September 20, 2002 to December 28, 2007}
\label{trend}
\end{figure}

At this point, it is worthwhile to provide historical context relating to the state of the financial markets in early 2008, at the same time when these reverse convertible notes were being sold to the public. Figure \ref{trend} was obtained from Google Finance's website and is based on historical closing prices from September 20, 2002 to December 28, 2007 of the S\&P 500 Financials Index.

As Figure \ref{trend} shows, the S\&P 500 Financial Index had undergone a sustained increase during the period 2002 to late 2007.  By early 2008, many investors were concerned that the bull market had gone too far and that stock prices might be due for substantial declines.  Therefore, it was reasonable to expect that $P(I_r < 0)$ was large for the $r$th quarter, hence, $P(I_r \ge 0)$ was small for that quarter.

Given the statistical evidence that momentum-trading strategies are successful over 3- to 12-month periods, we also would have expected $P(I_2 < 0 | I_1 < 0)$ to be high in late 2007, hence that $P(I_2 \ge 0 | I_1 < 0)$ was small.  Then it follows that
\begin{eqnarray*}
P(I_1 < 0, \hbox{ and } I_2 \ge 0) & \equiv & P(I_2 \ge 0 | I_1 < 0) \, 
P(I_1 < 0) \\
& < & P(I_2 \ge 0 | I_1 < 0),
\end{eqnarray*}
which proves that $P(I_1 < 0, \hbox{ and } I_2 \ge 0)$ also is small. By repeating this argument, we deduce that, in early 2008, each of the probabilities in the first six terms in (\ref{e(netpayment)}) are very small.  Therefore,
\begin{equation}
\label{e(np)2}
\begin{aligned}
E({\hbox{Net Payment}}) & \simeq 10 E(I_6|d_{\min} <-\tfrac 12 \hbox{ and } I_j < 0, 1 \leq j \leq 6)\\
                      &\quad \times P(d_{\min} <-\tfrac 12|I_j < 0, 1 \leq j \leq 6) \\
                      &\quad \times P(I_j < 0,  1 \leq j \leq 6)
\end{aligned}
\end{equation}
which, clearly, is negative.  Consequently, we deduce that clients who purchased these reverse convertible notes in early 2008  likely were destined to obtain negative net returns, on average.

 Even more can be deduced from (\ref{e(np)2}).  By late 2007,  $P(I_j <0, 1 \leq j \leq 6)$ and $P(d_{\min} <-\tfrac 12|I_j < 0,  1 \leq j \leq 6)$ both were very large.  Hence, by (\ref{e(np)2}),
\begin{equation}
\label{E(np)3}
E({\hbox{Net Payment}}) \simeq 10\times E(I_6|d_{\min} <-\tfrac 12\hbox{ and } I_j < 0,  1 \leq j \leq 6).
\end{equation}
If  it was believed in early 2008 that the financial markets were likely to have a precipitous, quarter-over-quarter, fall over the next 18 months, then it would mean that $I_6 \simeq d_{\min}$, in which case it follows from (\ref{E(np)3}) that
\begin{eqnarray*}
E({\hbox{Net Payment}}) & \simeq & 10 \times E(d_{\min} | d_{\min}< -\tfrac 12 \hbox{ and } I_j < 0, 1 \leq j \leq 6) \\
& < & 10 \times -0.5 \\
& = & -5.
\end{eqnarray*}
Consequently, the outcome of purchasing these reverse convertible notes in early 2008 was to undertake a substantial risk of {\it least} a 50\% loss of capital, on average.

Admittedly, the analysis provided above is based on a pessimistic view in 2007-2008 of coming trends in the financial markets.  To provide an analysis which is based on an unbiased approach toward future market trends, we will therefore treat the market as a ``random walk'' in which $I_1,\ldots, I_6$ are mutually independent and identically distributed random variables. These assumptions imply that medium-term momentum-based strategies will provide a trader with no advantages over the general market.  Also, such an assumption will be more profitable for a client who purchased reverse convertible notes in 2008, for it means that the various probabilities appearing in the first six cases in the expected net-payment function are likely to be more favorable to the clients than would have been the case under the real-world assumption underlying the inequality (\ref{realworld}).

Nevertheless, under the assumptions that $I_1,\ldots, I_6$ are mutually independent and identically distributed, we shall show that the expected net-payment to note-holders still remains significantly negative.  Hence, we will deduce that even in a ``random walk'' scenario an average note-holder would have faced losses except under highly optimistic environments.

Using the assumption that $I_1,\ldots, I_6$  are independent, we see that equation (\ref{e(netpayment)}) becomes
\begin{equation}
\label{e(netpayment)2}
\begin{aligned}
E(\hbox{Net Payment}) = \ &  0.52\times P( I_1 \ge 0)\\
       & + 1.04 \times P( I_1 < 0)\times P( I_2 \ge 0)\\
       & + 1.56 \times \left[ \prod_{j=1}^{2}{P( I_j < 0)}\right]\times P(I_3 \ge 0) \\
       & + 2.08 \times \left[ \prod_{j=1}^{3}{P( I_j < 0)} \right]\times P(I_4 \ge 0)\\
       & + 2.61\times \left[ \prod_{j=1}^{4}{P( I_j < 0)}\right]\times P( I_5 \ge 0 )\\
       & + 3.13 \times \left[\prod_{j=1}^{5}{P( I_j < 0)}\right]\times P( I_6 \ge 0 )\\
       & + 10  \times  E(I_6|d_{\min}< -\tfrac12 \hbox{ and } I_j < 0,  1 \leq j \leq 6)\\
       & \hspace{1 mm}\quad \quad \times P(d_{\min}< -\tfrac12|I_j < 0,  1 \leq j \leq 6) \times \prod_{j=1}^{6}{P( I_j < 0)}.
\end{aligned}
\end{equation}

Denote $ P(I_1 \geq 0)$ by $p$, because of the assumption that $I_1,\ldots, I_6$ are identically distributed, it follows that $p\equiv P(I_j \geq 0)$ for all $j=1, \ldots,6$. Note that, $p$ is the probability that the note is called, i.e. that the Index Return on any given observation date was non-negative. Then, we obtain 
\begin{equation}
\label{e(netpayment)3}
\begin{aligned}
                   E(\hbox{Net Payment})  =\ & 0.52p+ 1.04(1-p)p+1.56 (1-p)^2p\\ 
                      & + 2.08(1-p)^3p +2.61(1-p)^4p + 3.13(1-p)^5p\\
                      & +10 \times E(I_6|d_{\min}< -\tfrac12 \hbox{ and } I_j < 0,  1 \leq j \leq 6)\\
 & \quad \quad \times P(d_{\min}< -\tfrac12|I_j < 0, 1 \leq j \leq 6) \times (1-p)^6 .
\end{aligned}
\end{equation}

In order to derive explicit values for the terms 
$$E(I_6|d_{\min}< -\tfrac12 \hbox{ and } I_j < 0,  1 \leq j \leq 6)$$
and 
$$P(d_{\min}< -\tfrac12|I_j < 0, 1 \leq j \leq 6),$$
we would need to make strong assumptions about the random variables $I_1, \ldots, I_6$, which we prefer to avoid doing. Therefore, we will assign values to these terms to reflect a variety of market conditions, and then we will study the resulting behavior of the function (\ref{e(netpayment)3}) for each set of values. 

Define
\begin{equation}
\label{B1}
B_1 \equiv - E(I_6|d_{\min}< -\tfrac12 \hbox{ and } I_j < 0, 1\leq j \leq 6).
\end{equation}
Note that the expectation $E(I_6|d_{\min}< -\tfrac12 \hbox{ and } I_j < 0, 1\leq j \leq 6)$ clearly is negative; hence, $B_1 < 0$. Also, because $I_6 \ge -1$, then $ B_1 \leq 1$. Therefore, we have $0 < B_1 \le 1$.

Also, define
\begin{equation}
\label{B2}
B_2 \equiv P(d_{\min}< -\tfrac12|I_j < 0, 1\leq j\leq 6) ;
\end{equation}
clearly, $ 0 \le B_2 \leq 1$ because it is a probability. Moreover, by (\ref{Prob(B)}), $B_2 \neq 0$; therefore $0 < B_2 \le 1$. 

By (\ref{e(netpayment)3}), we obtain
\begin{equation}
\label{e(netpayment)41}
\begin{aligned}
            E(\hbox{Net Payment}) = \ & 0.52p+ 1.04(1-p)p+1.56 (1-p)^2p+ 2.08(1-p)^3p \\
                                                     & +2.61(1-p)^4p + 3.13(1-p)^5p- 10 B_1 B_2 (1-p)^6.
 \end{aligned}
\end{equation}
For $p =0$, we obtain $E(\hbox{Net Payment}) = -10B_1B_2 < 0$. Therefore, there will always be a small interval around $p=0$ where the expected net payment is negative.

As regards an interpretation for values of $B_1$, we offer the following. Suppose that $B_1$ is small; then it follows from (\ref{B1}) that $E(I_6|d_{\min}< -\tfrac12 \hbox{ and } I_j < 0, 1\leq j \leq 6)$ is close to zero {\it and} is negative. Hence, for small values of $B_1$, a note-holder will receive a small, but negative, overall expected return despite the conditioning event that $I_1,\ldots,I_6$ all are negative.  That is to say, although the index returns on all six observation dates were negative, and the minimum cumulative daily return is less than -50\%, the final outcome was that the average loss to note-holders nevertheless turned out to be small.

Hence, small values of $B_1$ indicate that the financial markets are likely to have maintained an optimistic outlook throughout the 18-month observation period despite consecutive quarterly losses.  This leads us to associate small values of $B_1$ with general optimism amongst market participants, and by a similar argument, we can associate large values of $B_1$ with broad market pessimism.

In the case of $B_2$, it follows from (\ref{B2}) that if $B_2$ is small then there is a small probability of any daily cumulative index return breaching the Trigger Level, conditional on six consecutive quarterly losses.  In a real-world situation, consecutive quarterly losses usually induces general gloom on market participants; so it would indeed be an optimistic assumption to believe that there remains a small probability of avoiding the Trigger Level.  Therefore, we see that small values of $B_2$ can be associated with a generally optimistic market outlook; and by the same argument, large values of $B_2$ can be associated with broad market pessimism.

\begin{figure}[H]
\centering
\includegraphics[scale=0.7]{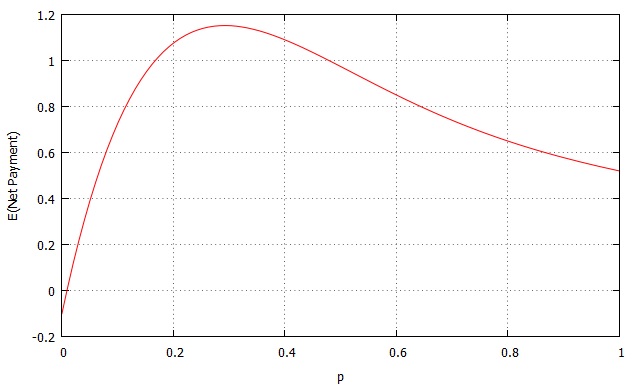}
\caption{A Graph of Expected Net Payments, where $B_1= 0.1$ and $B_2=0.1$}
\label{bound1}
\end{figure}

Consider the case in which $B_1= B_2=0.1$. Using these values, we plot (\ref{e(netpayment)41}) with respect to $p$, $0 \le p \le 1$, to obtain in Figure \ref{bound1} a graph of the expected net payment function:\\

In this case, the maximum expected net payment is 1.15, approximately, i.e., an average return of about 11.5\%. Also, the minimum expected net payment is $-0.1$, approximately, i.e., an average loss of about 10\%. Thus, even under this optimistic scenario, the average client could lose as much as 10\% of their capital.

We remark also that the expected net payment is maximized at $p = 0.29$, approximately.  Although it might have been presumed that consecutive quarterly increases in the financial markets, i.e., $p = 1$, would be more beneficial to note-holders, such a scenario would have caused the note to be called early due to the sustained market increases.  In such a case, paradoxically, average returns to note-holders would have been diminished.

Under a less-optimistic scenario for the financial markets, in which $B_1=B_2=0.5$, the graph of expected net payment function in (\ref{e(netpayment)41}) is given in Figure \ref{bound2}:

\begin{figure}[H]
\centering
\includegraphics[scale=0.7]{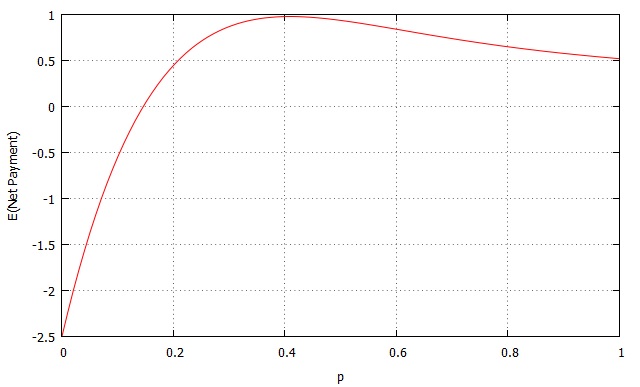}
\caption{A Graph of Expected Net Payments, where $B_1=0.5$ and $B_2=0.5$} 
\label{bound2}
\end{figure}

For this situation, the maximum expected net payment is 0.97, approximately, i.e., an average return of about 9.7\%. Also, the minimum expected net payment is $-2.5$, approximately, representing an average loss of about 25\% of principal. As in the previous scenario, moderate values of $p$ turn out to be more beneficial to note-holders than high values of $p$.

\begin{figure}[H]
\centering
\includegraphics[scale=0.7]{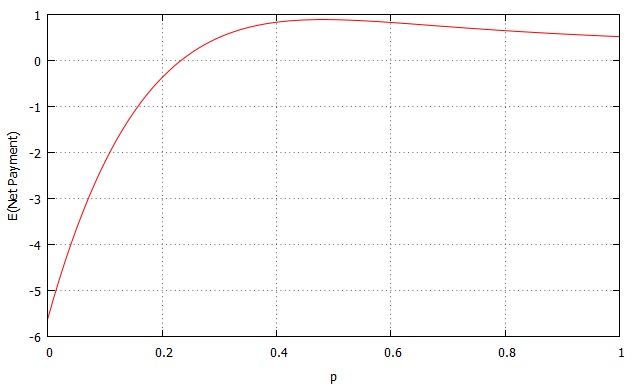}
\caption{A Graph of Expected Net Payments, where $B_1=0.7$ and $B_2=0.8$ }
\label{bound3}
\end{figure}

If financial markets are expected to undergo significant downturns where the values of $B_1$ and $B_2$ are relatively high, say $B_1=0.7$ and $B_2=0.8$, then we obtain in Figure \ref{bound3} the graph of the expected net payment function in (\ref{e(netpayment)41}): \\

Under this pessimistic scenario, the highest expected net payment is  0.89, approximately, which means that the maximum percentage return for the average note-holder was only about 8.9\%. On the other hand, the minimum expected net payment is $-5.6$, approximately, implying that the average note-holder could have suffered percentage losses of as much as 56\% of their capital.

We have studied up to now the exact values of the expected net payment function which, as we have seen, involves two unknown parameters, $B_1$ and $B_2$. We now derive upper bounds for the expected net payment function; these bounds will have the advantage of depending on only one unknown parameter. 

We know that $ d_{\min} \le I_r $ for any $r= 1, \dots,6$; therefore,
\begin{equation}
\label{case8_1}
P(d_{\min}< -\tfrac 12|I_j < 0, 1 \leq j \leq 6) \ge P(I_r< -\tfrac 12|I_j < 0, 1 \leq j \leq 6).
\end{equation}
Also, the expectation $E(I_6|d_{\min}< -\tfrac12 \hbox{ and } I_j < 0, 1\leq j \leq 6) < 0$ because $I_6 < 0 $ in the conditioning event. Therefore, by inequality (\ref{case8_1}), we see that
\begin{equation}
\label{case8_2}\begin {aligned}
 &E(I_6|d_{\min}< -\tfrac12 \hbox{ and } I_j < 0,  1 \leq j \leq 6) P(d_{\min}< -\tfrac12|I_j < 0, 1 \leq j \leq 6)\\
& \le E(I_6|d_{\min}< -\tfrac12 \hbox{ and } I_j < 0,  1 \leq j \leq 6) P(I_r< -\tfrac 12|I_j < 0, 1 \leq j \leq 6),
\end{aligned}
\end{equation}
and then, by (\ref{case8_2}), we obtain
\begin{equation}
\label{e(netpayment)4}
\begin{aligned}
E(\hbox{Net Payment}) \leq \ &  0.52p+ 1.04(1-p)p+1.56 (1-p)^2p\\ 
         & + 2.08(1-p)^3p +2.61(1-p)^4p + 3.13(1-p)^5p\\
         & +10(1-p)^6 E(I_6|d_{\min}< -\tfrac12 \hbox{ and } I_j < 0,  1 \leq j \leq 6)\\
 & \quad \quad \hspace{1 mm} \times P(I_r< -\tfrac 12|I_j < 0, 1 \leq j \leq 6).
\end{aligned}
\end{equation}

By the definition of conditional probability and the assumptions that $I_1, \ldots, I_6$ are independent and identically distributed,
\begin{equation}
\label{case8_3}
\begin{aligned}
P(I_r< -\tfrac 12|I_j < 0, 1 \leq j \leq 6)
&= P(I_r< -\tfrac 12|I_r< 0)\\
&=\frac{P(I_r< -\tfrac 12 {\hbox{ and }} I_r< 0)}{P(I_r< 0)}\\
&=\frac{P(I_r< -\tfrac 12)}{P(I_r< 0)}\\
&= (1-p)^{-1} P(I_r< -\tfrac 12).
\end{aligned}
\end{equation}
Thus, by (\ref{e(netpayment)4}) and (\ref{case8_3}), 
\begin{equation}
\label{e(netpayment)42}
\begin{aligned}
E(\hbox{Net } & \hbox{Payment})\\ \le & \ 0.52p+ 1.04(1-p)p+1.56 (1-p)^2p\\ 
   & + 2.08(1-p)^3p +2.61(1-p)^4p + 3.13(1-p)^5p\\
   & + 10 (1- p)^5 E(I_6|d_{\min}< -\tfrac 12 \hbox{ and } I_j < 0,  1 \leq j \leq 6)P(I_r< -\tfrac 12).
\end{aligned}
\end{equation}

We can also obtain a lower bound on $E(I_6|d_{\min}< -\tfrac 12 \hbox{ and } I_j < 0,  1 \leq j \leq 6).$  To that end, define 
$$
A_1= \{I_r< -\tfrac 12 \hbox{ and } I_j < 0, 1 \leq j \leq 6\}
$$
and
$$
A_2= \{ d_{\min} < -\tfrac 12 \hbox{ and } I_j < 0, 1 \leq j \leq 6\}.
$$
Because $d_{\min} \leq I_r$ for $r=1, \ldots,6$ then it follows that $A_1\subseteq A_2$.

Consider the cumulative index returns $d_1, \ldots,d_n$, viewed as continuous random variables. The sets $A_1$ and $A_2$ above clearly are events that depend on $d_1, \ldots,d_n$. Because $A_1\subseteq A_2$ then it follows that
\begin{equation}
\label{AvsB}
P(A_1) \leq P(A_2).
\end{equation}
We noted earlier in this section in (\ref{Prob(B)}) that $P(A_2) \neq 0$; by a similar argument, it follows that $P(A_1) \neq 0$. As mentioned previously, this condition holds because $I_1, \ldots, I_6$ are continuous random variables whose probability density function is strictly positive on the interval (-1,0), a condition which is equivalent to the assumption that there is a non-zero probability of the markets moving downward in any quarter. 

Let $f(d_1,\ldots,d_n)$ denote the joint probability density function of $I_1,\ldots,I_6$. Since $I_6 \equiv I_6(d_1,\ldots,d_n)$ is a function of the random variables $d_1,\ldots,d_n$ then, by the definition of expected value, we obtain
\begin{equation}
\label{expectation}
\begin{aligned}
E(I_6|d_{\min}< -\tfrac 12 & \hbox{ and } I_j < 0, 1 \leq j \leq 6) \\ 
& \equiv E(I_6|A_2)\\
& = \frac{1}{P(A_2)}\int_{A_2} I_6(x_1,\ldots,x_n) \cdot f(x_1,\ldots,x_n) \dd x_1 \cdots \dd x_n.
\end{aligned}
\end{equation}
Because $I_6 < 0$ for all  $(x_1,\ldots,x_n) \in A_2$ and because $A_1 \subseteq A_2$ then it follows that 
\begin{multline}
\label{int}
\int_{A_2} I_6(x_1,\ldots,x_n) \cdot f(x_1,\ldots,x_n) \dd x_1 \cdots \dd x_n \\ 
\leq \int_{A_1} I_6(x_1,\ldots,x_n) \cdot f(x_1,\ldots,x_n) \dd x_1 \cdots \dd x_n.
\end{multline}
It now follows that from (\ref{AvsB}), (\ref{expectation}) and (\ref{int}) that
\begin{equation}
\label{exp2}
\begin{aligned}
E(I_6|A_2) & =\frac{1}{P(A_2)}\int_{A_2} I_6(x_1,\ldots,x_n) \cdot f(x_1,\ldots,x_n) \dd x_1 \cdots \dd x_n \\
& \leq \frac{1}{P(A_1)}\int_{A_1} I_6(x_1,\ldots,x_n) \cdot f(x_1,\ldots,x_n) \dd x_1 \cdots \dd x_n\\
& \equiv E(I_6|A_1).
\end{aligned}
\end{equation}
This proves that
\begin{equation}
\label{compare_exp}
 E(I_6|d_{\min}< -\tfrac 12 \hbox{ and } I_j < 0,  1 \leq j \leq 6) 
 \leq E(I_6|I_r< -\tfrac 12 \hbox{ and } I_j < 0,  1 \leq j \leq 6). 
\end{equation}
Hence for any $r=1,\ldots,6$,  
\begin{equation}
\label{e(netpayment)5}
\begin{aligned}
E(\hbox{Net } & \hbox{Payment})\\ \le & \ 0.52p+ 1.04(1-p)p+1.56 (1-p)^2p\\ 
                &+ 2.08(1-p)^3p +2.61(1-p)^4p + 3.13(1-p)^5p\\
                &+10 (1- p)^5 E(I_6|I_r < -\tfrac 12 \hbox{ and } I_j < 0,  1 \leq j \leq 6)P(I_r< -\tfrac 12).
\end{aligned}
\end{equation}
Also, note that the inequality (\ref{e(netpayment)5}) involves the random variables $I_1,\ldots,I_6$  only, and does not depend on $d_{\min}$.  Thus,  we can select $r=6$ to obtain
\begin{equation}
\label{e(netpayment)6}
\begin{aligned}
E(\hbox{Net } & \hbox{Payment})\\ \le & \ 0.52p+ 1.04(1-p)p+1.56 (1-p)^2p\\ 
                      & + 2.08(1-p)^3p +2.61(1-p)^4p + 3.13(1-p)^5p\\
                      & +10 (1- p)^5 E(I_6|I_6 < -\tfrac 12\hbox{ and } I_j < 0,  1 \leq j \leq 6)P(I_6< -\tfrac 12).
\end{aligned}
\end{equation}
It is clear that 
\begin{equation}
\label{EofI6bound}
E(I_6|I_6 < -\tfrac 12\hbox{ and } I_j < 0,  1 \leq j \leq 6) < -\tfrac 12.
\end{equation}
We now introduce the notation $\tau = P(I_6 < - \tfrac 12)$; then it clear that $p +\tau \leq 1$. Therefore, it follows from (\ref{e(netpayment)6}) and (\ref{EofI6bound}) that
\begin{equation}
\label{e(netpayment)7}
\begin{aligned}
E(\hbox{Net Payment}) \le & \ 0.52p+ 1.04(1-p)p+1.56 (1-p)^2p\\ 
                      & + 2.08(1-p)^3p +2.61(1-p)^4p + 3.13(1-p)^5p\\
                      & -5 (1- p)^5 \tau.
\end{aligned}
\end{equation}

\begin{figure}[H]
\centering
\includegraphics[scale=0.7]{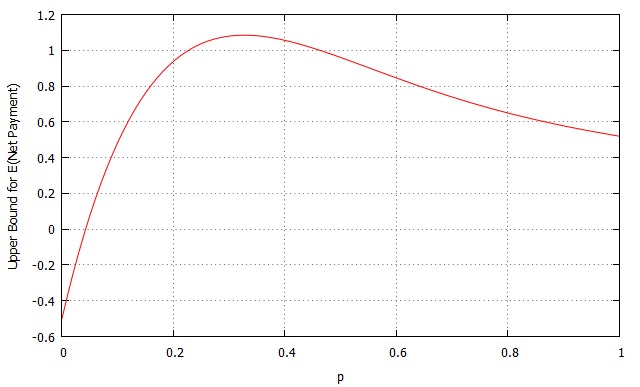}[p]
\caption{ Upper Bound for Expected Net Payments, where $P(I_6<-\tfrac 12) =0.1$}
\label{bound4}
\end{figure}

Under an optimistic scenario, we expect there is a small probability of $I_6$ breaching the Trigger Level, i.e., $\tau$ is small. For instance, let $\tau=0.1 $, then we obtain in Figure \ref{bound4} a graph of the right-hand side of (\ref{e(netpayment)7}):\\

In this case, an upper bound for the maximum expected net payment is 1.08, approximately, resulting in an upper bound for the average return of about $10.8\%$. On the other hand, an upper bound for the minimum expected net payment is $-0.5$, approximately, i.e., an average loss of about 5\%. 

\begin{figure}[H]
\centering
\includegraphics[scale=0.7]{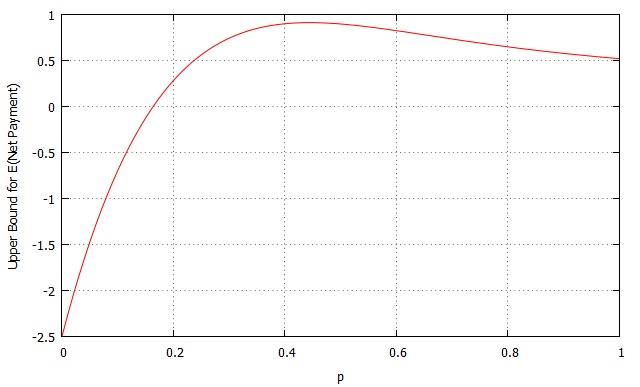}
\caption{Upper Bound for Expected Net Payments, where $P(I_6<-\tfrac 12) =0.5$}
\label{bound5}
\end{figure}

Suppose we let $\tau=0.5$; then we obtain in Figure \ref{bound5} a graph of the upper bound for the expected net payment function in (\ref{e(netpayment)7}):\\

Under this mildly pessimistic scenario, the highest value for an upper bound for expected net payment is  0.91, approximately; and an upper bound for the lowest expected net payment is $-2.5$, approximately. Equivalently, an upper bound for the average return to note-holders is about 9.1\%, whereas an upper bound for the average {\it loss} to note-holders is about 25\%. 

Now consider a highly pessimistic scenario for the financial markets in which $\tau=0.8$; then Figure \ref{bound6} depicts an upper bound for the expected net payment given by (\ref{e(netpayment)7}):

\begin{figure}[H]
\centering
\includegraphics[scale=0.7]{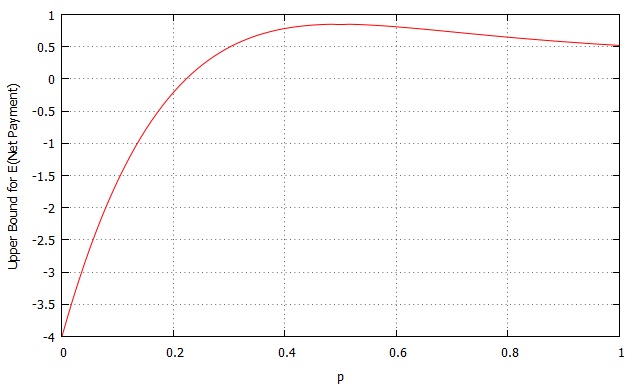} 
\caption{Upper Bound for Expected Net Payments, where $P(I_6<-\tfrac 12) =0.8$}
\label{bound6}
\end{figure}

For this scenario, an upper bound for the maximum expected net payment is  0.85, approximately. On the other hand an upper bound for the lowest expected net payment is  $-4$, approximately. These values imply that an upper bound for the the return for the average client is about 8.5\% while an upper bound for the possible loss is about 40\%. 

In conclusion, Figures 3.3--3.8 depict upper bounds for expected net payments under various scenarios. As a general rule, the highest values for expected net payments occur at moderate values of $p$ and not at high values of $p$, as one would first presume. Under a scenario in which markets are expected to undergo significant upward trends and the value of $p$ is high, the note will be called early and therefore will limit returns to note-holders. 

Based on our analysis of the six scenarios depicted in Figures 3.3--3.8, the best-case scenario leads to a possible maximum an average return of 11.5\%, approximately, and a possible average loss of 10\%, approximately. The worst-case scenario is an average return of 8\%,  approximately, and an average loss of  56\%, approximately. 

Bearing in mind that only six scenarios were described above in detail, we also provide a more comprehensive analysis by graphing the upper bound function in (\ref{e(netpayment)7})  over all values of $(p,\tau)$ such that $ 0 \leq p , \tau \leq 1$ and $ p + \tau  \leq 1$. Then we obtain the following three-dimensional plot of expected net payments:\\ 

\begin{figure}[H]
\centering
\includegraphics[scale=0.7]{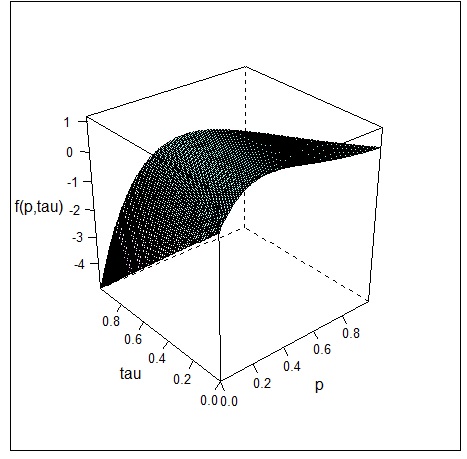} 
\caption{Upper Bound for Expected Net Payments as a function of $p$ and $\tau$ }
\label{np_qp}
\end{figure}

From this analysis, the best-case scenario is an expected net payment of at most 1.16, approximately, at $p=0.28$ and $\tau=0$. On the other hand, the worst-case scenario results in an expected net payment of at most $-5$, approximately, at  $p=0$ and $\tau=1$. Thus, the net percentage return for note-holders ranges from $-50\%$ to 11.6\%, approximately. This range implies that on average, the possible returns to note-holders are substantially lower than the potential losses. As a result, an average note-holder would have suffered a significant loss of capital.  

\section{A Second Interpretation of the Payment Procedure}
\label{sec:secondinterpretation}
\setcounter{equation}{0}

We now provide a second interpretation of the payment procedure and analyze it  similarly to the way in which the previous interpretation was assessed. In the interpretation of the payment procedure studied in the previous section, if an Index Ending Level breached the Trigger Level during any quarter of the observation period but the Index Return was positive on a subsequent observation date, then the possibility of a positive return still existed. Because the prospectus was unclear on this issue, we will assume in this section that if an Index Ending Level breached the Trigger Level on any trading day during a quarter, i.e. if at least one cumulative daily index return was less than $-50\%$, then the client is destined to receive a negative return at the Final Valuation date. 

Throughout this section the notation for the six observation dates and the daily cumulative index return will remain the same as before. We also need to define a collection of minimum daily cumulative index returns as a function of the various observation dates, viz., \\

$d_{\min,1}$ : Minimum cumulative daily index return on May 5, 2008 

$d_{\min,2}$ : Minimum cumulative daily index return on August 5, 2008

$d_{\min,3}$ : Minimum cumulative daily index return on November 5, 2008

$d_{\min,4}$ : Minimum cumulative daily index return on February 5, 2009

$d_{\min,5}$ : Minimum cumulative daily index return on May 5, 2009

$d_{\min,6}$ : Minimum cumulative daily index return on  August 5, 2009
 \\

The last minimum cumulative daily index return, $d_{\min:6}$, is equivalent to $d_{\min}$ in the analysis provided in Section \ref{sec:firstinterpretation}. Also, from this notation it follows that:
\begin{equation}
\label{dmin}
d_{\min,1} \geq d_{\min,2}\geq d_{\min,3}\geq d_{\min,4}\geq d_{\min,5}\geq d_{\min,6} .
\end{equation}

With this notation, the steps for this interpretation of the payment procedure can now be described as follows:\\

\begin{itemize}
\item[ ] Step 1:\vspace{-7mm} \begin{adjustwidth}{1.5cm}{1.5cm} Calculate the Index Return, $I_r$ on the observation date for the $r$th quarter.
\end{adjustwidth}

\item[ ] Step 2: \vspace{-7mm} \begin{adjustwidth}{1.5cm}{1.5cm} If the Index Return on the observation date is non-negative, i.e. if $I_r\geq 0$, then the note is called and the client receives an amount as stated in the pricing supplement on the Final Valuation date.
\end{adjustwidth}

\item[ ] Step 3: \vspace{-7mm} \begin{adjustwidth}{1.5cm}{1.5cm}  Calculate the cumulative returns, $d_i$, for each trading date during the quarter corresponding to the given observation date and determine the minimum cumulative daily index return, $d_{\min,r}$ at this observation date.
\end{adjustwidth}

\item[ ] Step 4: \vspace{-7mm} \begin{adjustwidth}{1.5cm}{1.5cm} If the Index Return on the observation date is negative and the minimum cumulative daily return is greater than $-50\%$, i.e. if $I_r < 0$ and $d_{\min,r} \geq -50\%$, then we return to Step 1 for the next observation date and repeat the process.
\end{adjustwidth}

\item[ ] Step 5: \vspace{-7mm} \begin{adjustwidth}{1.5cm}{1.5cm}If the Index Return on the observation date is negative and minimum cumulative daily return is less than $-50\% $, i.e. if $I_r < 0$ and $d_{\min,t} <-50\%$, then no further action is taken until the Final Valuation date.
\end{adjustwidth}
\end{itemize}
\bigskip

Steps 1-4 are carried out on every observation date until the Final Valuation date or until the note is called. 

We remark that on the Final Valuation date if the note has not been called then all previous minimum cumulative daily index returns are greater than $-50\%$ and by equation (\ref{dmin}) $d_{\min, 5}  \ge -50\%$.\\

At the Final Valuation date, one of the following three cases can occur:\\

\begin{itemize}
\item[ ] Case 1:\vspace{-7mm} \begin{adjustwidth}{1.5cm}{1.5cm}  If the Index Return on the observation date is non-negative  i.e. $I_6\geq 0$, then the note is called and the client receives an amount as stated in the pricing supplement on the Final Valuation date.
\end{adjustwidth}
\item[ ] Case 2: \vspace{-7mm} \begin{adjustwidth}{1.5cm}{1.5cm} If the Index Return is negative and the minimum cumulative daily index return is greater than $-50\%$ on the Final Valuation date, i.e. if $I_6 < 0$ and $d_{\min,6} \geq -50\%$, then the client receives \$10 at maturity.
\end{adjustwidth}
\item[ ] Case 3:\vspace{-7mm} \begin{adjustwidth}{1.5cm}{1.5cm}  If any minimum cumulative daily index return is less than $-50\%$, i.e., if $d_{\min,r} < -50\%$ for at least one $r=1, \ldots,6$, then the client receives a reduced payment equal to $10(1+I_6)$, representing a negative return of  $I_6$ to note-holders.
\end{adjustwidth}
\end{itemize}
\bigskip

Note that on the final valuation date that if at least one $d_{\min,r} < -50\%$ then it follows from inequality (\ref{dmin}) that 
$$
d_{\min,6} \leq 50\%.
$$

The following tree diagram describes this interpretation of the payment procedure:

\begin{figure}[H]
\hspace{-2in}
\includegraphics{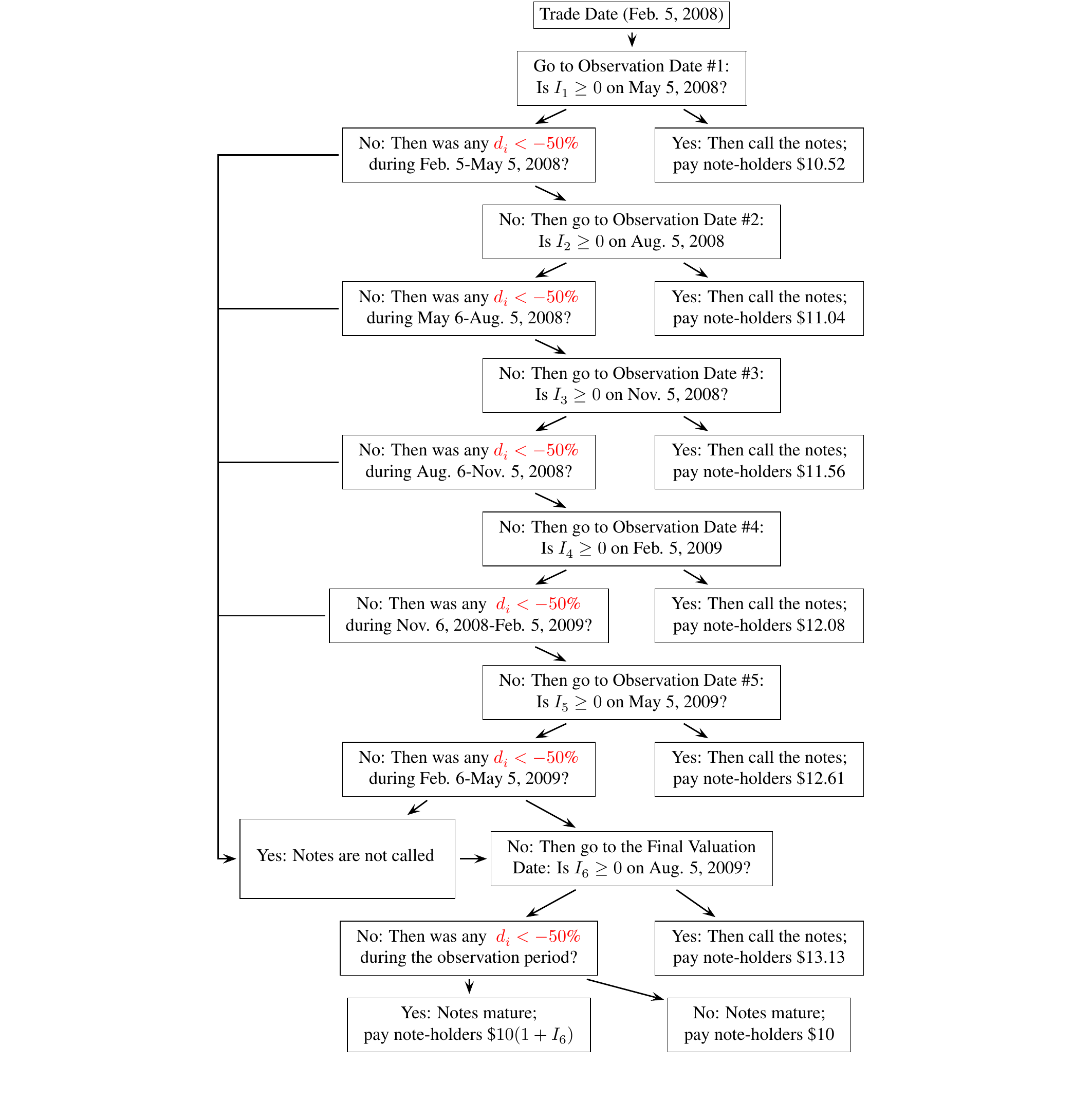}
\vskip-20pt
\caption{Schematic Description of a Payment Procedure}
\label{payment2}
\end{figure}


For this interpretation of the payment procedure, we apply the information provided in the pricing supplement to deduce that the net payment in US dollars is given by: 

\begin{equation}
\label{np_fun}
\hbox{Net Payment} = 
\begin{cases}
0.52, \hbox{ if } I_1 \ge 0\\
1.04, \hbox{ if } I_1 < 0 \hbox{ and }  I_2 \geq 0 \hbox{ and } d_{\min,1} \geq -\tfrac 12\\
1.56,  \hbox{ if } I_j < 0, j=1,2 \hbox{ and }  I_3 \geq 0 \hbox{ and } d_{\min,2} \geq -\tfrac 12\\
2.08, \hbox{ if } I_j < 0, j=1,2,3 \hbox{ and }  I_4 \geq 0 \hbox{ and } d_{\min,3} \geq -\tfrac 12\\
2.61, \hbox{ if } I_j < 0, 1 \leq j \leq 4 \hbox{ and }  I_5 \geq 0 \hbox{ and } d_{\min,4} \geq -\tfrac 12\\
3.13, \hbox{ if } I_j < 0, 1 \leq j \leq 5 \hbox{ and }  I_6 \geq 0 \hbox{ and } d_{\min,5} \geq -\tfrac 12\\
0,  \hspace{5mm} \hbox{ if } I_j < 0, 1 \leq j \leq 6 \hbox{ and } d_{\min,6} \geq -\tfrac 12\\
10 I_6,   \hbox{ if } I_j < 0, 1 \leq j \leq 6 \hbox{ and }  d_{\min,6} < -\tfrac 12
\end{cases} 
\end{equation}

\begin{itemize}
\item The first six cases occur if any of the index returns, $I_1, \ldots, I_6$ are non-negative and the minimum cumulative daily index return of the previous quarter is at least $-50\%$ i.e., $I_r \ge 0$ and $ d_{\min,r-1} \geq -50\%$, in which case, the note is called on the corresponding observation date, and the client receives a positive return in these instances.

\item The seventh case occurs if all of the index returns, $I_1, \ldots, I_6$ are negative and every minimum cumulative daily  index return is at least $-50\% $. In this case, the note matures and the client receives a net payment of zero, i.e., the client  breaks even.

\item The eighth and last case occurs if any $ d_{\min,r} \leq -50\%$ , i.e. the minimum cumulative daily index return during some quarter of the observation period fell below $-50\%$. In this case, the note matures and the client receives a negative net payment of  $10I_6$.
\end{itemize}
Based on our function for net payment above, we obtain:
\begin{equation}
\label{E(Netpayment)4.1}
\begin{aligned}
E(\hbox{Net Payment}) = \ & 0.52 \times P( I_1 \ge 0)\\
                           & + 1.04 \times P(I_1< 0 \hbox{ and }  I_2 \geq 0 \hbox{ and } d_{\min,1} \geq -\tfrac 12) \\
                           &+ 1.56 \times P( I_j < 0, j=1,2 \hbox{ and }  I_3 \geq 0 \hbox{ and } d_{\min,2} \geq -\tfrac 12  ) \\
                      & + 2.08 \times P( I_j < 0,1 \le j \le 3 \hbox{ and }  I_4 \geq 0 \hbox{ and } d_{\min,3} \geq -\tfrac 12)\\
                     & +2.61 \times P( I_j < 0, 1 \le j \le 4 \hbox{ and }  I_5 \geq 0 \hbox{ and } d_{\min,4} \geq -\tfrac 12)\\
                  & +3.13 \times P(  I_j < 0, 1\le j \le 5 \hbox{ and }  I_6 \geq 0 \hbox{ and } d_{\min,5} \geq -\tfrac 12 )\\
  & + 0 \times P(I_j < 0, 1 \le j \le 6 \hbox{ and }  d_{\min,6}\geq -\tfrac 12) \\
                 & + 10 \times E(I_6|d_{\min,6}< -\tfrac 12 \hbox{ and } I_j < 0, 1 \leq j \leq 6)\\
                     & \quad\quad \times P(d_{\min,6}< -\tfrac 12|I_j < 0,  1 \leq j \leq 6) \\
       & \quad \quad \times P(I_j < 0,  1 \le j \le 6).
\end{aligned}
\end{equation}

Consider the probabilities appearing in the second thru sixth terms in (\ref{E(Netpayment)4.1}), It is clear that for each $r = 1,\ldots,6$,
\begin{multline}
\label{prob_compare}
P(I_j<0,j=1,\ldots,r-1, I_r \ge 0,  d_{\min,r-1} \ge \tfrac 12) \\ \leq P(I_j<0,j=1,\ldots,r-1, I_r \ge 0 ).
\end{multline}

Also, as noted earlier, $d_{\min,6} \equiv d_{\min}$, the minimum cumulative index return encountered in Section \ref{sec:firstinterpretation}. Therefore, it follows from (\ref{E(Netpayment)4.1}) and (\ref{prob_compare}) that
\begin{equation}
\label{E(Netpayment)4.2}
\begin{aligned}
E(\hbox{Net Payment}) \leq\ & 0.52 \times P( I_1 \ge 0)\\
                  &+ 1.04\times P(I_1< 0, \hbox{ and }  I_2 \geq 0) \\
                  &+ 1.56 \times P( I_j < 0, j=1,2 \hbox{ and } I_3 \geq 0) \\
                  & + 2.08 \times P( I_j < 0, j=1,2,3 \hbox{ and } I_4 \geq 0)\\
                  & +2.61 \times P( I_j < 0, 1 \leq j \leq 4 \hbox{ and } I_5 \geq 0)\\
                  & +3.13 \times P(  I_j < 0, 1 \leq j \leq 5 \hbox{ and } I_6 \geq 0)\\
  & + 0 \times P(I_j < 0, 1 \leq j \leq 6 \hbox{ and } d_{\min}\geq -\tfrac 12) \\
          & + 10\times E(I_6|d_{\min} <-\tfrac 12 \hbox{ and } I_j < 0,  1 \leq j \leq 6)\\
                  &\quad \quad \times P(d_{\min} <-\tfrac 12|I_j < 0, 1 \leq j \leq 6) \\
                  &\quad \quad \times P(I_j < 0, 1 \leq j \leq 6).
\end{aligned}
\end{equation}

The expression on the right hand side is precisely the expected net payment function which we derived in Section \ref{sec:firstinterpretation}. Therefore, the average return to clients from the interpretation of the payment procedure considered here in Section \ref{sec:secondinterpretation} can be no greater than the average return from the interpretation in Section \ref{sec:firstinterpretation}. Likewise, average losses for this interpretation will be greater than losses in the interpretation of the payment procedure in Section \ref{sec:firstinterpretation}. Consequently, it is clear that clients are likely to suffer even greater losses under the new interpretation of the payment procedure considered in this latter section.

\section{Conclusions}
\label{conclusions}
\setcounter{equation}{0}

In our analysis, we did not consider the effects of income taxes or commissions related to the  sales of these reverse convertible notes. The inclusion of these expenses would have resulted in smaller returns and greater losses for note-holders. 

Further, any financial adviser who was required to exercise fiduciary duty to their clients, could not have done so by recommending that clients purchase these notes.  The SEC has stated that a financial adviser has a fiduciary duty to make reasonable {\emph{investment}} recommendations to clients. Graham \cite{Graham73} defines an investment operation as ``one which upon thorough analysis promises safety of principal and an adequate return.'' Had financial advisers carried out a {\emph{thorough analysis}} of these reverse convertible notes then they would have realized that the notes promised neither {\emph{safety of principal}} nor {\emph{adequate return}} to clients. As a result, any financial adviser acting as a fiduciary, should have opposed the purchase of these notes.

We have seen recently an increase in the sales of similar notes by many financial firms; the prospectuses for these notes are available on the SEC's website. This increase in sales perhaps is due to the fact that the financial markets have become increasingly bullish since 2009 and as a result these notes now appear more attractive to persons looking to purchase fixed-income securities with yields higher than available through U.S. Treasury securities. However, we predict that if  markets undergo substantial downward movements then clients are again likely to lose substantial capital from purchases of these notes. 

Further, these notes were issued during a time when future market conditions were uncertain and direct sales of stock may accelerated further downturns. Thus, these notes were a clever marketing scheme used by financial institutions which wanted to insure themselves against substantial market declines.

It is particularly noteworthy to observe the actual outcome for the reverse convertible note \cite{Lehman08} which was analyzed in the previous sections.  In this case, the Index Starting Level was $369.44$, the Trigger Level was $184.72$, and the Index Ending Level and outcome for each observation date are provided in Table \ref{actualoutcome}:

\bigskip

\begin{table}[!h]
\caption{Actual Outcome of a Reverse Convertible Note}
\label{actualoutcome}
\centering
\begin{tabular}{lll}
  &  & \\
Observation Date  & Index Ending Level & Outcome \\
\hline
May 5, 2008 & 365.48 & Below Index Starting Level and  Above \\
& & Trigger Level; Securities NOT called  \\
\\
August 5, 2008 &  302.05 & Below Index Starting Level  and Above \\
 & & Trigger Level; Securities NOT called  \\
\\
November 5, 2008 &201.77 & Below Index Starting Level  Above \\
& & Trigger Level; Securities NOT called  \\
\\
 Feb 5, 2008 &121.51 & Below Index Starting Level and Below \\
& & Trigger Level;  Securities NOT called  \\
\\
 May 5, 2009 & 155.52 & Below Index Starting Level and Below \\
& & Trigger Level; Securities NOT called \\
\\
 August 5, 2009 &189.37 & Below Index Starting Level  and Above \\
& & Trigger Level; Securities NOT called  \\
\\
 {\bf{Settlement Amount}}  & \bf{{(per \$10)}}  & {\bf{\$5.13  (total return of {\color{red}{-48.7\%}}, or a}} \\
& & {\bf{compound return of {\color{red}{-35.9\%}} per}} \\
& & {\bf{annum)}} \\
\hline
\end{tabular}
\end{table}

\bigskip
\medskip

\noindent 
Note that the index return on the Final Valuation date equals
\begin{eqnarray*}
\frac{\hbox{Index Starting Return} - \hbox{Index Ending Level}}{\hbox{Index Starting Level}} & = & \frac{189.37 -
 369.44 }{369.44}\\  
& = & -0.487, 
\end{eqnarray*}
which represents a total loss to the note-holder of 48.7\%. Thus, the settlement amount is
\begin{eqnarray*}
\$10 \times (1 + \hbox{Index Return on the final valuation date}) &=& \$10 \times (1 - 0.487)\\ & =& \$ 5.13.
\end{eqnarray*}


%


\end{document}